\documentclass[a4paper,11pt]{article}

\usepackage[dvips]{color,graphicx}

\definecolor{red  }{rgb}{1,0,0}
\definecolor{blue }{rgb}{0,0,1}
\definecolor{green}{rgb}{0,1,0}

\usepackage{graphicx,amssymb,amsmath,bm,latexsym}
\newcommand{\vs}[1]{\vspace{#1 mm}}

\newcommand{\la}{\lambda}
\newcommand{\pa}{\partial}

\begin{document}

\begin{titlepage}

\vskip .5in

\begin{center}

{\Large\bf Super $w_{\infty}$ 3-algebra } \vskip .5in

{\large Min-Ru Chen$^{a,b}$,  Ke Wu$^{a,} $\footnote{
wuke@mail.cnu.edu.cn} and Wei-Zhong Zhao$^{a,c,}$\footnote{zhaowz100@163.com}
}\\
\vs{10}
$^a${\em School of Mathematical Sciences, Capital Normal University,
Beijing 100048, China} \\
$^b${\em College of Mathematics and Information Sciences, Henan
University, Kaifeng 475001, China} \\
$^c${\em Institute of Mathematics and Interdisciplinary Science,
Capital
Normal University, Beijing 100048, China } \\

\vskip .2in \vspace{.3in}

\begin{abstract}

We investigate the super high-order Virasoro 3-algebra. By applying
the appropriate  scaling limits on the generators, we obtain the
super $w_{\infty}$ 3-algebra which satisfies the generalized
fundamental identity condition. We also define a super Nambu-Poisson
bracket which satisfies the generalized skewsymmetry, Leibniz rule
and  fundamental identity. By means of this super Nambu-Poisson
bracket, the realization of the super $w_{\infty}$ 3-algebra is
presented.

\end{abstract}

\end{center}



\vfill

\end{titlepage}

\section{Introduction}

Since Nambu \cite{N} first proposed 3-bracket for the generalized
Hamiltonian dynamics, 3-algebras have  attracted much interest from
physical and mathematical points of view. Recently with the
development of string theory, it is found that 3-algebras have the
important applications in M-theory, such as multiple M2-branes
\cite{BL1}-\cite{Gus} and  Nambu-Poisson M5-brane theory
\cite{HM}-\cite{CFHT}. 3-algebras can be realized in two ways, i.e.,
classical Nambu and quantal brackets. These two brackets are defined
by means of  multi-variable Jacobians and
 antisymmetrized products of three linear operators, respectively.
 The properties of various 3-algebras have been widely investigated \cite{CFHT}-\cite{AI}.
The Virasoro algebra is an infinite-dimensional algebra
which plays an important role in conformal field theory. The
centerless Virasoro 3-algebra, i.e., Virasoro-Witt 3-algebra has
been constructed in the literature \cite{CFZ}-\cite{CFJMZ}. The
$W_{\infty}$ algebra is the higher-spin extensions of the Virasoro
algebra \cite{PRS}. Quite recently by applying a double scaling
limits on the generators of the $W_{\infty}$ algebra, Chakrabortty
et al.\cite{CKJ} constructed a $w_{\infty}$ 3-algebra which
satisfies the fundamental identity (FI) condition  of 3-algebra.

The supersymmetric generalizations of 3-algebras are of general
interest \cite{Pa}-\cite{SS2}. Recently Sakakibara \cite{Sa}
constructed the super Nambu-Poisson algebra and demonstrated its
connection with the generalized Batalin-Vilkovisky algebra. Soroka
et al. proposed the Grassmann-odd Nambu brackets and investigated
their properties in their serial paper \cite{SS1}\cite{SS2}. The
supersymmetric generalizations of the Virasoro and $w$ algebras have
been well investigated. As to their super 3-algebras, to our best
knowledge, it has not been reported so far in the existing
literature. The propose of this paper is to present the super
$w_{\infty}$ 3-algebra which satisfies the generalized FI condition.

This paper is organized as follows. In the next section, we
investigate the super high-order Virasoro (SHOV) algebra and its
3-algebra. Then by applying an appropriate  scaling limits on the
generators of the SHOV algebra, we give the super $w_{\infty}$
3-algebra. In section 3, we define the super Nambu-Poisson bracket
to realize the super $w_{\infty}$ 3-algebra. We end this paper with
the concluding remarks in section 4.

\section{SHOV algebra and super $w_{\infty}$ 3-algebra}

The generators of SHOV algebra are given by
\begin{eqnarray}\label{eq:gen}
&&L_m^i=(-1)^i{\lambda}^{i-\frac{1}{2}} z^{m+i}\partial_z^i, \nonumber\\
&&\bar L_m^i=(-1)^{i}{\la}^{i+\frac{3}{2}} z^{m+i}\theta\partial_\theta\partial_z^i,\nonumber\\
&&h_r^{\alpha+\frac{1}{2}}=(-1)^{\alpha +1}\la^{\alpha+\frac{1}{2}} z^{r
+\alpha}\partial_\theta\partial_z^\alpha,\nonumber\\
&&\bar h_r^{\alpha+\frac{1}{2}}=(-1)^{\alpha+1}
\la^{\alpha+\frac{1}{2}} z^{r+\alpha}\theta\partial_z^\alpha,
\end{eqnarray}
where $ i, \alpha\in Z_+$, $m, r\in Z$ and $\lambda$ is an arbitrary
parameter.

Their communication relations are
\begin{eqnarray}\label{eq:shov}
[L_m^i, L_n^j]=\sum_{p=0}^{i} (-1)^p\la^{p-\frac{1}{2}}C_i^p
B_p^{n+j} L_{m+n}^{i+j-p}
 -\sum_{p=0}^j(-1)^p\la^{p-\frac{1}{2}}
C_j^p B_p^{m+i}L_{m+n}^{i+j-p},\nonumber
\end{eqnarray}
\begin{eqnarray}
[L_m^i, \bar L_n^j]=\sum_{p=0}^{i}(-1)^p\la^{p-\frac{1}{2}} C_i^p
B_p^{n+j} \bar L_{m+n}^{i+j-p}
 -\sum_{p=0}^j(-1)^p\la^{p-\frac{1}{2}}C_j^p B_p^{m+i}
 \bar L_{m+n}^{i+j-p},\nonumber
\end{eqnarray}
\begin{eqnarray}
[L_m^i, h_r^{\alpha+\frac{1}{2}}]=\sum_{p=0}^{i}(-1)^p
\la^{p-\frac{1}{2}} C_i^p B_p^{r+\alpha}
h_{m+r}^{\alpha+i-p+\frac{1}{2}}
 -\sum_{p=0}^\alpha(-1)^p
\la^{p-\frac{1}{2}}C_\alpha^p B_p^{m+i}
h_{m+r}^{\alpha+i-p+\frac{1}{2}},\nonumber
\end{eqnarray}
\begin{eqnarray}
[L_m^i, \bar h_r^{\alpha+\frac{1}{2}}]=\sum_{p=0}^{i}(-1)^p
\la^{p-\frac{1}{2}} C_i^p B_p^{r+\alpha} \bar
h_{m+r}^{\alpha+i-p+\frac{1}{2}}
 -\sum_{p=0}^\alpha(-1)^p
\la^{p-\frac{1}{2}}C_\alpha^p B_p^{m+i} \bar
h_{m+r}^{\alpha+i-p+\frac{1}{2}},\nonumber
\end{eqnarray}
\begin{eqnarray}
[\bar L_m^i, h_r^{\alpha+\frac{1}{2}}]=-\sum_{p=0}^{\alpha}
(-1)^{p}\la^{p+\frac{3}{2}} C_\alpha^p
B_p^{m+i}h_{m+r}^{\alpha+i-p+\frac{1}{2}},\nonumber
\end{eqnarray}
\begin{eqnarray}
[\bar L_m^i, \bar L_n^j]=\sum_{p=0}^{i} (-1)^{p}\la^{p+\frac{3}{2}}
C_i^p B_p^{n+j}  \bar L_{m+n}^{i+j-p}
 -\sum_{p=0}^j(-1)^{p}
\la^{p+\frac{3}{2}}C_j^p B_p^{m+i} \bar L_{m+n}^{i+j-p},\nonumber
\end{eqnarray}
\begin{eqnarray}
[\bar L_m^i, \bar h_r^{\alpha+\frac{1}{2}}]=\sum_{p=0}^{i}(-1)^{p}
\la^{p+\frac{3}{2}} C_i^p B_p^{r+\alpha} \bar
h_{m+r}^{\alpha+i-p+\frac{1}{2}},\nonumber
\end{eqnarray}
\begin{eqnarray}
[h_r^{\alpha+\frac{1}{2}}, \bar
h_s^{\beta+\frac{1}{2}}]&=&\sum_{p=0}^{\alpha}
(-1)^{p}\la^{p+\frac{3}{2}} C_\alpha^p B_p^{s+\beta}
L_{r+s}^{\alpha+\beta-p} +\sum_{p=0}^{\beta}(-1)^p
\la^{p-\frac{1}{2}}
C_\beta^pB_p^{r+\alpha}\bar L_{r+s}^{\alpha+\beta-p}\nonumber\\
&-&\sum_{p=0}^{\alpha}(-1)^p \la^{p-\frac{1}{2}}
C_\alpha^pB_p^{s+\beta} \bar L_{r+s}^{\alpha+\beta-p},
\end{eqnarray}
where
$B_p^n=\left\{\begin{array}{cc}
n(n-1)\cdots(n-p+1),& p\leq n\\
0,                  &p>n,\end{array}\right. $
$C_n^m=\frac{n(n-1)\cdots(n-m+1)}{m!}$
and the communication relation is defined by
\begin{eqnarray}\label{eq:gbracket}
[f,g]=fg-(-1)^{|f||g|}gf,
\end{eqnarray}
$|f|$ and $|g|$ are the parity of $f$ and $g$, respectively. A
notational convention used frequently in the rest of this paper is
that for any arbitrary h, the symbol $|h|$ appearing in the exponent
of $(-1)$ is to be understood as the parity of h. When $\lambda=1$,
(\ref{eq:shov}) leads to the communication relations of SHOV algebra
derived by Zha and Zhao \cite{zz}. Recently Chakrabortty et al.
\cite{CKJ} constructed a $w_{\infty}$ 3-algebra by applying a double
scaling limits on the generators of $W_{\infty}$ algebra. In order
to construct the super $w_{\infty}$ 3-algebra, we introduce a
parameter $\lambda$ into the generators of SHOV algebra
(\ref{eq:gen}).  Not as done by Chakrabortty et al., we'll construct
the super $w_{\infty}$ 3-algebra by taking a single scaling limit on
the generators (\ref{eq:gen}). Therefore the parameter $\lambda$
plays a crucial role in the following investigation.

Let us define a super 3-bracket as follows:
\begin{eqnarray}\label{eq:bracket1}
[f, g, h]=[f, g]h+(-1)^{|f|(|g|+|h|)}[g, h]f+(-1)^{|h|(|f|+|g|)}[h, f]g,
\end{eqnarray}
where the commutator $[\ , \ ]$ is defined by (\ref{eq:gbracket}).
Substituting the  generators (\ref{eq:gen}) into the generalized ternary commutator (\ref{eq:bracket1}),
we may obtain the SHOV 3-algebra.
Due to too many 3-algebra relations, we only list some of them that will be used in the late discussion.
\begin{eqnarray}\label{eq:lim1}
[L_m^i, L_n^j, L_k^h] &=&[(\sum_{p=0}^{i} C_i^p
B_p^{n+j}-\sum_{p=0}^j C_j^p B_p^{m+i})\sum_{q=0}^{i+j-p}
C_{i+j-p}^q B_q^{k+h}
 +(\sum_{p=0}^{j} C_j^p B_p^{k+h}\nonumber\\
 &&-\sum_{p=0}^h C_h^p B_p^{n+j})\sum_{q=0}^{j+h-p} C_{j+h-p}^q B_q^{m+i}
 +(\sum_{p=0}^{h} C_h^p B_p^{m+i}\nonumber\\
 &&-\sum_{p=0}^i C_i^p B_p^{k+h})\sum_{q=0}^{h+i-p} C_{h+i-p}^q B_q^{n+j}]
 (-1)^{p+q}\la^{p+q-1} L_{m+n+k}^{i+j+h-p-q},\nonumber
\end{eqnarray}
\begin{eqnarray}
[L_m^i, L_n^j, \bar{L}_k^h]
&=&[(\sum_{p=0}^{i} C_i^p B_p^{n+j}-\sum_{p=0}^j C_j^p
B_p^{m+i})\sum_{q=0}^{i+j-p} C_{i+j-p}^q B_q^{k+h}
 +(\sum_{p=0}^{j} C_j^p B_p^{k+h}\nonumber\\
 &&-\sum_{p=0}^h C_h^p B_p^{n+j})\sum_{q=0}^{j+h-p} C_{j+h-p}^q B_q^{m+i}
 +(\sum_{p=0}^{h} C_h^p B_p^{m+i}\nonumber\\
 &&-\sum_{p=0}^i C_i^p B_p^{k+h})\sum_{q=0}^{h+i-p} C_{h+i-p}^q B_q^{n+j}](-1)^{p+q}
 \la^{p+q-1}\bar{L}_{m+n+k}^{i+j+h-p-q},\nonumber
\end{eqnarray}
\begin{eqnarray}
[L_m^i, L_n^j, h_r^{\alpha+\frac{1}{2}}]
&=&[(\sum_{p=0}^{i} C_i^p B_p^{n+j}-\sum_{p=0}^j C_j^p
B_p^{m+i})\sum_{q=0}^{i+j-p} C_{i+j-p}^q B_q^{r+\alpha}
 +(\sum_{p=0}^{j} C_j^p B_p^{r+\alpha}\nonumber\\
 &&-\sum_{p=0}^\alpha C_\alpha^p B_p^{n+j})\sum_{q=0}^{j+\alpha-p} C_{j+\alpha-p}^q B_q^{m+i}
 +(\sum_{p=0}^{\alpha} C_{\alpha}^p B_p^{m+i}\nonumber\\
 &&-\sum_{p=0}^i C_i^p B_p^{r+\alpha})\sum_{q=0}^{\alpha+i-p}
 C_{\alpha+i-p}^q B_q^{n+j}](-1)^{p+q}\la^{p+q-1} h_{m+n+r}^{i+j+\alpha-p-q+\frac{1}{2}},\nonumber
\end{eqnarray}
\begin{eqnarray}
[L_m^i, L_n^j, \bar{h}_r^{\alpha+\frac{1}{2}}]
&=&[(\sum_{p=0}^{i} C_i^p B_p^{n+j}-\sum_{p=0}^j C_j^p
B_p^{m+i})\sum_{q=0}^{i+j-p} C_{i+j-p}^q B_q^{r+\alpha}
 +(\sum_{p=0}^{j} C_j^p B_p^{r+\alpha}\nonumber\\
 &&-\sum_{p=0}^\alpha C_\alpha^p B_p^{n+j})\sum_{q=0}^{j+\alpha-p} C_{j+\alpha-p}^q B_q^{m+i}
 +(\sum_{p=0}^{\alpha} C_{\alpha}^p B_p^{m+i}\nonumber\\
 &&-\sum_{p=0}^i C_i^p B_p^{r+\alpha})
 \sum_{q=0}^{\alpha+i-p} C_{\alpha+i-p}^q B_q^{n+j}]
 (-1)^{p+q}\la^{p+q-1} \bar{h}_{m+n+r}^{i+j+\alpha-p-q+\frac{1}{2}},\nonumber
\end{eqnarray}
\begin{eqnarray}
[L_m^i, h_r^{\alpha+\frac{1}{2}}, \bar{h}_s^{\beta+\frac{1}{2}}]
&=&[(\sum_{p=0}^{i} C_i^p B_p^{r+\alpha}
 -\sum_{p=0}^\alpha C_\alpha^p B_p^{m+i})\sum_{q=0}^{i+\alpha-p}C_{i+\alpha-p}^q
 B_q^{s+\beta}\nonumber\\
 &+&\sum_{p=0}^{\alpha}\sum_{q=0}^{\alpha+\beta-p} C_\alpha^p
B_p^{s+\beta}C_{\alpha+\beta-p}^qB_q^{m+i}](-1)^{p+q}\la^{p+q+1}
L_{m+r+s}^{i+\alpha+\beta-p-q}\nonumber\\
&+&[-(\sum_{p=0}^{i} C_i^p B_p^{r+\alpha}
 -\sum_{p=0}^\alpha C_\alpha^p B_p^{m+i})\sum_{q=0}^{i+\alpha-p}C_{i+\alpha-p}^q
 B_q^{s+\beta}
 +(\sum_{p=0}^{\beta}C_\beta^pB_p^{r+\alpha}\nonumber\\
&-&\sum_{p=0}^{\alpha}C_\alpha^pB_p^{s+\beta})\sum_{q=0}^{\alpha+\beta-p}C_{\alpha+\beta-p}^qB_q^{m+i}
-(\sum_{p=0}^\beta C_\beta^p B_p^{m+i}\nonumber\\
 &-&\sum_{p=0}^{i} C_i^p
B_p^{s+\beta})\sum_{q=0}^{i+\beta-p}C_{i+\beta-p}^qB_q^{r+\alpha}]
(-1)^{p+q} \la^{p+q-1}\bar{L}_{m+r+s}^{i+\alpha+\beta-p-q}.
\end{eqnarray}

Let us take the scaling limit $\la \rightarrow 0 $ and for
convenience denote  the generators with the same notations for this
and other kinds of limit throughout this paper, then from
(\ref{eq:lim1}), we obtain the following super $w_{\infty}$
3-algebra:
\begin{eqnarray}\label{eq:3alg1}
[L_m^i,L_n^j,L_k^h]&=&(h(n-m)+j(m-k)+i(k-n))L_{m+n+k}^{i+j+h-1},\nonumber
\end{eqnarray}
\begin{eqnarray}
[L_m^i, L_n^j, \bar L_k^h]&=&-[L_m^i, \bar L_k^h,  L_n^j]=[\bar L_k^h,  L_m^i, L_n^j]\nonumber\\
&=&(h(n-m)+j(m-k)+i(k-n))\bar L_{m+n+k}^{i+j+h-1},\nonumber
 \end{eqnarray}
\begin{eqnarray}
[L_m^i, L_n^j, h_r^{\alpha+\frac{1}{2}}]&=&-[L_m^i,  h_r^{\alpha+\frac{1}{2}}, L_n^j]=
[h_r^{\alpha+\frac{1}{2}}, L_m^i, L_n^j]\nonumber\\
&=&(\alpha
(n-m)+j(m-r)+i(r-n))h_{m+n+r}^{i+j+\alpha-1+\frac{1}{2}},\nonumber
\end{eqnarray}
\begin{eqnarray}
[L_m^i, L_n^j, \bar h_r^{\alpha+\frac{1}{2}}]&=&-[L_m^i, \bar h_r^{\alpha+\frac{1}{2}},  L_n^j]=
[\bar h_r^{\alpha+\frac{1}{2}}, L_m^i, L_n^j]\nonumber\\
&=&(\alpha (n-m)+j(m-r)+i(r-n))\bar
h_{m+n+r}^{i+j+\alpha-1+\frac{1}{2}},\nonumber
\end{eqnarray}
\begin{eqnarray}
[L_m^i, h_r^{\alpha+\frac{1}{2}}, \bar h_s^{\beta+\frac{1}{2}}]&=&
[L_m^i, \bar{h}_s^{\beta+\frac{1}{2}}, h_r^{\alpha+\frac{1}{2}}]
=[h_r^{\alpha+\frac{1}{2}}, \bar{h}_s^{\beta+\frac{1}{2}},L_m^i]
=[\bar{h}_s^{\beta+\frac{1}{2}}, h_r^{\alpha+\frac{1}{2}}, L_m^i]\nonumber\\
&=&-[h_r^{\alpha+\frac{1}{2}}, L_m^i, \bar{h}_s^{\beta+\frac{1}{2}}]
=-[\bar{h}_s^{\beta+\frac{1}{2}}, L_m^i, h_r^{\alpha+\frac{1}{2}}]\nonumber\\
&=&(i(r-s)+\alpha (s-m)+\beta(m-r))\bar
L_{m+r+s}^{i+\alpha+\beta-1},
\end{eqnarray}
with all other 3-brackets vanishing. It should be pointed out that
under this kind of scaling limit we have the non-null 3-algebra
(\ref{eq:3alg1}), but (\ref{eq:shov}) becomes the null algebra. Note
that the first 3-algebraic relation in (\ref{eq:3alg1}) is the
so-called $w_{\infty}$ 3-algebra derived in Ref.\cite{CKJ}. It is
known that this $w_{\infty}$ 3-algebra satisfies the usually FI
condition. As to the super $w_{\infty}$ 3-algebra (\ref{eq:3alg1}),
it should satisfy the generalized FI condition due to the involution
of fermionic generators. We find  that the super $w_{\infty}$
3-algebra (\ref{eq:3alg1}) satisfies the following generalized FI
condition:
\begin{eqnarray}\label{eq:FI}
[A, B, [C, D, E]]&=&[[A, B, C], D, E]+(-1)^{(|A|+|B|)|C|}[C, [A, B, D], E]\nonumber\\
&+&(-1)^{(|A|+|B|)(|C|+|D|)}[C, D, [A, B, E]].
\end{eqnarray}
When the generators in (\ref{eq:FI}) are bosonic, (\ref{eq:FI}) reduces to the usually FI condition of 3-algebra.

\section{Super Nambu-Poisson bracket}
For the graded communicating bracket (\ref{eq:gbracket}), it is well-known that the classical limit is given by
\begin{eqnarray}\label{eq:lim23}
\{\ , \ \}=\lim\limits_{\hbar \rightarrow 0 }\frac{1}{{\bf
i}\hbar}[\ ,\ ].
\end{eqnarray}
Let us take the generators of SHOV algebra as follows:
\begin{eqnarray}\label{eq:gen2}
&&L_m^i={(-{\bf i}\hbar)}^iz^{m+i}\partial_z^i,\ \ \ \ \ \ \ \ \ \ \ \ \ \
\bar L_m^i={(-{\bf i}\hbar)}^{i+2}z^{m+i}\theta\partial_\theta\partial_z^i,\nonumber\\
&&h_r^{\alpha+\frac{1}{2}}={(-{\bf i}\hbar)}^{\alpha+1}z^{r+\alpha}\partial_\theta\partial_z^\alpha,\ \ \ \
\bar h_r^{\alpha+\frac{1}{2}}={(-{\bf i}\hbar)}^{\alpha+1}z^{r+\alpha}\theta\partial_z^\alpha,
\end{eqnarray}
where ${\bf i}=\sqrt{-1}$ and  $\hbar$ is introduced in the above
generators. Substituting the generators (\ref{eq:gen2}) into
(\ref{eq:gbracket}) and taking the limit (\ref{eq:lim23}), we obtain
the classical super $w_{\infty}$ algebra
\begin{eqnarray}
&&\{L_m^i, L_n^j\}=(mj-ni)L_{m+n}^{i+j-1},\nonumber\\
&&\{L_m^i, \bar L_n^j\}=(mj-ni)\bar L_{m+n}^{i+j-1},\nonumber\\
&&\{L_m^i, h_r^{\alpha+\frac{1}{2}}\}=(m\alpha-ri)h_{m+r}^{i+\alpha-\frac{1}{2}},\nonumber\\
&&\{L_m^i, \bar h_r^{\alpha+\frac{1}{2}}\}=(m\alpha-ri)\bar h_{m+r}^{i+\alpha-\frac{1}{2}},\nonumber\\
&&\{\bar L_m^i, h_r^{\alpha+\frac{1}{2}}\}=\{\bar L_m^i, \bar
h_r^{\alpha+\frac{1}{2}}\}=
\{\bar L_m^i, \bar L_n^j \}=0,\nonumber\\
&&\{h_r^{\alpha+\frac{1}{2}}, \bar
h_s^{\beta+\frac{1}{2}}\}=(s\alpha- r\beta)\bar
L_{r+s}^{\alpha+\beta-1}.
\end{eqnarray}
This super $w_{\infty}$ algebra is different with ones derived by
Pope and Shen \cite{ps}.

Let us turn to discuss the case of 3-bracket. For the 3-bracket defined by (\ref{eq:bracket1}),
we require  the following classical limit to be hold \cite{T}:
\begin{eqnarray}\label{eq:lim33}
\{\  ,\  ,\ \}=\lim\limits_{\hbar \rightarrow 0 }\frac{1}{{\bf
i}\hbar}[\ , \ , \ ],
\end{eqnarray}
where $\{\  ,\  ,\ \}$ should be understood as the super Nambu-Poisson bracket.

Substituting the generators (\ref{eq:gen2}) into (\ref{eq:bracket1})
and taking the classical limit (\ref{eq:lim33}), we obtain the
classical super $w_{\infty}$ 3-algebra
\begin{eqnarray}\label{eq:3alg2}
\{L_m^i,L_n^j,L_k^h\}&=&(h(n-m)+j(m-k)+i(k-n))L_{m+n+k}^{i+j+h-1},\nonumber
\end{eqnarray}
\begin{eqnarray}
\{L_m^i, L_n^j, \bar L_k^h\}&=&-\{L_m^i, \bar L_k^h,  L_n^j\}=\{\bar L_k^h,  L_m^i, L_n^j\}\nonumber\\
&=&(h(n-m)+j(m-k)+i(k-n))\bar L_{m+n+k}^{i+j+h-1},\nonumber
 \end{eqnarray}
\begin{eqnarray}
\{L_m^i, L_n^j, h_r^{\alpha+\frac{1}{2}}\}&=&-\{L_m^i,  h_r^{\alpha+\frac{1}{2}}, L_n^j\}=
\{h_r^{\alpha+\frac{1}{2}}, L_m^i, L_n^j\}\nonumber\\
&=&(\alpha
(n-m)+j(m-r)+i(r-n))h_{m+n+r}^{i+j+\alpha-1+\frac{1}{2}},\nonumber
\end{eqnarray}
\begin{eqnarray}
\{L_m^i, L_n^j, \bar h_r^{\alpha+\frac{1}{2}}\}&=&-\{L_m^i, \bar h_r^{\alpha+\frac{1}{2}},  L_n^j\}=
\{\bar h_r^{\alpha+\frac{1}{2}}, L_m^i, L_n^j\}\nonumber\\
&=&(\alpha (n-m)+j(m-r)+i(r-n))\bar
h_{m+n+r}^{i+j+\alpha-1+\frac{1}{2}},\nonumber
\end{eqnarray}
\begin{eqnarray}
\{L_m^i, h_r^{\alpha+\frac{1}{2}}, \bar h_s^{\beta+\frac{1}{2}}\}&=&
\{L_m^i, \bar{h}_s^{\beta+\frac{1}{2}}, h_r^{\alpha+\frac{1}{2}}\}
=\{h_r^{\alpha+\frac{1}{2}}, \bar{h}_s^{\beta+\frac{1}{2}},L_m^i\}
=\{\bar{h}_s^{\beta+\frac{1}{2}}, h_r^{\alpha+\frac{1}{2}}, L_m^i\}\nonumber\\
&=&-\{h_r^{\alpha+\frac{1}{2}}, L_m^i, \bar{h}_s^{\beta+\frac{1}{2}}\}
=-\{\bar{h}_s^{\beta+\frac{1}{2}}, L_m^i, h_r^{\alpha+\frac{1}{2}}\}\nonumber\\
&=&(i(r-s)+\alpha (s-m)+\beta(m-r))\bar
L_{m+r+s}^{i+\alpha+\beta-1},
\end{eqnarray}
with all other 3-brackets vanishing, which exactly match with
(\ref{eq:3alg1}).

The concept of quantization plays a fundamental role in quantum
physics. It is well-known that the limit relation (\ref{eq:lim23})
establishes the relation between classical and quantum mechanics.
The limit relation (\ref{eq:lim33}) can be regarded as the analog of
(\ref{eq:lim23}).  Takhtajan \cite{T} investigated the quantization
of Nambu mechanics based on (\ref{eq:lim33}).  In the previous
section, we pointed out that by applying the scaling limit $\la
\rightarrow 0 $ on the generators of SHOV algebra (\ref{eq:gen}), we
have the null algebra and non-null super $w_{\infty}$ 3-algebra.
When we introduce $\hbar$  in the generators of SHOV algebra, under
the limits (\ref{eq:lim23}) and (\ref{eq:lim33}), we find that the
non-null super $w_{\infty}$ algebra and 3-algebra can be obtained,
respectively.

To present the realization of the classical super $w_{\infty}$
3-algebra (\ref{eq:3alg2}), we define the following super
Nambu-Poisson bracket:
\begin{eqnarray}\label{eq:bracket2}
\{f,g,h\}
&=&\frac{\pa f}{\pa x_1}\frac{\pa g}{\pa x_2}\frac{\pa h}{\pa x_3}
-\frac{\pa f}{\pa x_1}\frac{\pa g}{\pa x_3}\frac{\pa h}{\pa
x_2}+\frac{\pa f}{\pa x_2}\frac{\pa g}{\pa x_3}\frac{\pa h}{\pa
x_1}\nonumber\\
&-&\frac{\pa f}{\pa x_2}\frac{\pa g}{\pa x_1}\frac{\pa h}{\pa
x_3}+\frac{\pa f}{\pa x_3}\frac{\pa g}{\pa x_1}\frac{\pa h}{\pa
x_2}-\frac{\pa f}{\pa x_3}\frac{\pa g}{\pa x_2}\frac{\pa h}{\pa
x_1}\nonumber\\
&=&\sum_\sigma(-1)^{\varepsilon(\sigma)}\frac{\pa f}{\pa
x_{\sigma(1)}}\frac{\pa g}{\pa x_{\sigma(2)}}\frac{\pa h}{\pa x_{\sigma(3)}},
\end{eqnarray}
where $f$, $g$ and $h$ are the functions of the bosonic variables
$x_i, i=1,2,3$ and Grassmann variables $\theta_1$ and $\theta_2$,
$\sigma\in Symm(3)$,  $Symm(3)$ is a symmetric group of 3 elements
and $\varepsilon(\sigma)$ is the parity of permutation $\sigma$. It
should be pointed out that the derivative term  with respect to the
Grassmann variables does not be introduced in (\ref{eq:bracket2}).

It is known that the bosonic Nambu bracket is characterized by the three properties, i.e.,
 skewsymmetry, the Leibniz rule  and the fundamental identity.
 For the super Nambu-Poisson bracket (\ref{eq:bracket2}), these three properties should be generalized.
According to (\ref{eq:bracket2}), it is easy to prove that the
following generalized skew-symmetric properties are satisfied:
\begin{eqnarray}
&&\{g,f,h\}=(-1)^{1+|f||g|}\{f,g,h\},\nonumber\\
&&\{f,h,g\}=(-1)^{1+|g||h|}\{f,g,h\},\nonumber\\
&&\{h,g,f\}=(-1)^{1+|f||g|+|f||h|+|g||h|}\{f,g,h\}.
\end{eqnarray}

Let us consider the 3-bracket $\{f_1f_2,g,h\}$. By means of  (\ref{eq:bracket2}), we obtain
\begin{eqnarray}
\{f_1f_2, g, h\}&=&\sum_\sigma(-1)^{\varepsilon(\sigma)}(\frac{\partial f_1}{\partial
x_{\sigma(1)}}f_2+f_1\frac{\partial f_2}{\partial
x_{\sigma(1)}})\frac{\partial g}{\partial x_{\sigma(2)}}\frac{\partial h}{\partial x_{\sigma(3)}}\nonumber\\
&=&\sum_\sigma(-1)^{\varepsilon(\sigma)}(-1)^{|f_1||f_2|}f_2\frac{\partial
f_1}{\partial x_{\sigma(1)}}\frac{\partial g}{\partial x_{\sigma(2)}}\frac{\partial
h}{\partial
x_{\sigma(3)}}\nonumber\\
&+&\sum_\sigma(-1)^{\varepsilon(\sigma)}f_1\frac{\partial
f_2}{\partial
x_{\sigma(1)}}\frac{\partial g}{\partial x_{\sigma(2)}}\frac{\partial h}{\partial x_{\sigma(3)}}\nonumber\\
&=&f_1\{f_2, g, h\}+(-1)^{|f_1||f_2|}f_2\{f_1, g, h\}.
\end{eqnarray}
In a similar way, we have another two Leibniz rule relations
\begin{eqnarray}
\{f,g_1g_2,h\}&=&(-1)^{|g_2|(|f|+|g_1|)}g_2\{f,g_1,h\}+(-1)^{|g_1||f|}g_1\{f,g_2,h\}\nonumber\\
&=&(-1)^{|g_2||h|}\{f,g_1,h\}g_2+(-1)^{|g_1||f|}g_1\{f,g_2,h\},
\end{eqnarray}
\begin{eqnarray}
\{f,g,h_1h_2\}&=&\{f,g,h_1\}h_2+(-1)^{|h_1|(|f|+|g|)}h_1\{f,g,h_2\}\nonumber\\
&=&\{f,g,h_1\}h_2+(-1)^{|h_1||h_2|}\{f,g,h_2\}h_1.
\end{eqnarray}

To discuss the fundamental identity, we first give the following relations
by means of  (\ref{eq:bracket2}):
\begin{eqnarray}\label{eq:pfi1}
\{A, B, \{C, D, E\}\}&=&\{A, B, \sum_{\sigma^{\prime}}(-1)^{\varepsilon(\sigma^{\prime})}\frac{\pa
C}{\pa x_{\sigma^{\prime}(1)}}\frac{\pa D}{\pa x_{\sigma^{\prime}(2)}}\frac{\pa E}{\pa x_{\sigma^{\prime}(3)}}\}\nonumber\\
&=&\sum_{\sigma,\sigma^\prime}(-1)^{\varepsilon(\sigma)+\varepsilon(\sigma^\prime)}
\frac{\pa A}{\pa x_{\sigma(1)}}\frac{\pa B}{\pa x_{\sigma(2)}}
(\frac{\pa^2C }{\pa x_{\sigma^{\prime}(1)}\pa
x_{\sigma(3)}}\frac{\pa D}{\pa x_{\sigma^{\prime}(2)}}\frac{\pa
E}{\pa x_{\sigma^{\prime}(3)}} \nonumber\\
&+&\frac{\pa C }{\pa x_{\sigma^{\prime}(1)}}\frac{\pa^2 D}{\pa x_{\sigma^{\prime}(2)}\pa
x_{\sigma(3)}}\frac{\pa E}{\pa
x_{\sigma^{\prime}(3)}}
+\frac{\pa C }{\pa
x_{\sigma^{\prime}(1)}}\frac{\pa D}{\pa x_{\sigma^{\prime}(2)}}\frac{\pa^2 E}{\pa x_{\sigma^{\prime}(3)}\pa x_{\sigma(3)}}),
\end{eqnarray}
\begin{eqnarray}\label{eq:pfi2}
\{\{A, B, C\}, D, E\}
&=&\sum_{\sigma,\sigma^\prime}(-1)^{\varepsilon(\sigma)+\varepsilon(\sigma^\prime)}
(\frac{\pa^2 A}{\pa x_{\sigma(1)}\pa x_{\sigma^\prime(1)}}\frac{\pa
B}{\pa x_{\sigma(2)}}\frac{\pa C }{\pa x_{\sigma(3)}}
+\frac{\pa A}{\pa x_{\sigma(1)}}\frac{\pa^2 B}{\pa x_{\sigma(2)}\pa
x_{\sigma^\prime(1)}}\nonumber\\
&&\frac{\pa C }{\pa x_{\sigma(3)}}
+\frac{\pa A}{\pa x_{\sigma(1)}}\frac{\pa B}{\pa
x_{\sigma(2)}}\frac{\pa^2 C }{\pa x_{\sigma(3)}\pa
x_{\sigma^\prime(1)}})\frac{\pa D}{\pa
x_{\sigma^\prime(2)}}\frac{\pa E}{\pa
x_{\sigma^\prime(3)}},
\end{eqnarray}
\begin{eqnarray}\label{eq:pfi3}
\{C, \{A, B, D\}, E\}&=&\sum_{\sigma,\sigma^\prime}(-1)^{\varepsilon(\sigma)+\varepsilon(\sigma^\prime)}
\frac{\pa C}{\pa x_{\sigma^\prime(1)}}(\frac{\pa^2 A}{\pa
x_{\sigma(1)}\pa x_{\sigma^\prime(2)}}\frac{\pa B}{\pa
x_{\sigma(2)}} \frac{\pa D}{\pa x_{\sigma(3)}}
+\frac{\pa A}{\pa
x_{\sigma(1)}}\nonumber\\
&&\frac{\pa^2 B}{\pa x_{\sigma(2)}\pa
x_{\sigma^\prime(2)}} \frac{\pa D}{\pa x_{\sigma(3)}}
+\frac{\pa A}{\pa x_{\sigma(1)}}\frac{\pa B}{\pa x_{\sigma(2)}}
\frac{\pa^2 D}{\pa x_{\sigma(3)}\pa x_{\sigma^\prime(2)}})\frac{\pa
E}{\pa x_{\sigma^\prime(3)}},
\end{eqnarray}
\begin{eqnarray}\label{eq:pfi4}
\{C, D, \{A, B, E\}\}
&=&\sum_{\sigma,\sigma^\prime}(-1)^{\varepsilon(\sigma)+\varepsilon(\sigma^\prime)}
\frac{\partial C}{\partial x_{\sigma^\prime(1)}}\frac{\partial D}{\partial
x_{\sigma^\prime(2)}} (\frac{\partial^2 A}{\partial x_{\sigma(1)}\partial
x_{\sigma^\prime(3)}}\frac{\partial B}{\partial x_{\sigma(2)}} \frac{\partial
E}{\partial x_{\sigma(3)}}\nonumber\\
&&+\frac{\partial A}{\partial x_{\sigma(1)}}\frac{\partial^2
B}{\partial x_{\sigma(2)}\partial x_{\sigma^\prime(3)}} \frac{\partial
E}{\partial x_{\sigma(3)}}
+\frac{\partial A}{\partial x_{\sigma(1)}}\frac{\partial B}{\partial x_{\sigma(2)}}
\frac{\partial^2 E}{\partial x_{\sigma(3)}\partial x_{\sigma^\prime(3)}}).
\end{eqnarray}
Substituting  equations (\ref{eq:pfi1})-(\ref{eq:pfi4}) into
(\ref{eq:bracket1}), we find that the super Nambu-Poisson bracket
(\ref{eq:bracket2}) satisfies the generalized FI condition
(\ref{eq:FI}) with the substitution $[\ , \ , \ ]\rightarrow \{\  ,\
,\ \}$.

Let us take
\begin{eqnarray}\label{eq:gen3}
&&L_m^i=\sqrt{z}\exp[(i-\frac{1}{2})x-2my],\nonumber\\
&&\bar L_m^i=\theta_1\theta_2\sqrt{z}\exp[(i-\frac{1}{2})x-2my],\nonumber\\
&&h_r^{\alpha+\frac{1}{2}}=\theta_1\sqrt{z}\exp[(\alpha-\frac{1}{2})x-2ry],\nonumber\\
&&\bar
h_r^{\alpha+\frac{1}{2}}=-\theta_2\sqrt{z}\exp[(\alpha-\frac{1}{2})x-2ry].
\end{eqnarray}
It is noticed from (\ref{eq:gen3}) that
 $L_m^i$  is  the generator of $w_{\infty}$ 3-algebra presented in \cite{CKJ}.
Substituting the  generators (\ref{eq:gen3}) into
(\ref{eq:bracket2}), we obtain the classical super $w_{\infty}$
3-algebra (\ref{eq:3alg2}).

\vskip 20cm
\section{Concluding Remarks}

We have investigated the SHOV 3-algebra. By applying the appropriate
scaling limits on the generators, we obtained the super $w_{\infty}$
3-algebra.  We also found that this super $w_{\infty}$ 3-algebra
satisfies the generalized FI condition. To present the realization
of the  super $w_{\infty}$ 3-algebra, we defined a  super
Nambu-Poisson bracket which satisfies the generalized skewsymmetry,
 Leibniz rule  and  fundamental identity, but  the derivative term with respect to
the Grassmann variables are not involved  in the 3-bracket.
Moreover, we tried to introduce the Grassmann derivative term in the
super Nambu-Poisson bracket such that the super $w_{\infty}$
3-algebra can be realized. Unfortunately, we did not succeed in
finding any one. Whether there exists this kind of super
Nambu-Poisson bracket still deserves further study. Furthermore the
application of this super $w_{\infty}$ 3-algebra in physics should
be of interest.

\section*{Acknowledgements}

 This work is partially supported
by NSF projects (10975102, 10871135 and 11031005).


\end{document}